\documentclass{skaox2006}
\usepackage{graphicx}
\begin{document}
   \title{Local Universe Science with the E-ELT}

   \author{Michael R. Merrifield}

   \institute{School of Physics \& Astronomy, University of
     Nottingham, University Park, Nottingham, NG7 2RD, UK
             }

             \abstract{This paper briefly reviews some of the exciting
               studies of the local Universe that will be enabled with
               the European Extremely Large Telescope (E-ELT).  As
               illustrative examples, it summarizes a few of the
               scientific goals that have been set for this instrument
               for studies of young starburst clusters, evolved stars,
               the Galactic centre, Galactic structure,
               nucleo-chronometry, the interpretation of the ``Spite
               Plateau,'' and the properties of resolved stellar
               populations in external galaxies.  It finishes on a note
               of warning that we really need to be even more
               innovative and adventurous if we are to justify the
               cost of the E-ELT.}
   \maketitle
%
%
\section{Introduction}

There is not a great deal of point in trying to reproduce the full
local Universe science case for the European Extremely Large Telescope
(E-ELT) within these few pages for two reasons.  First, there already
exist far more comprehensive documents that seek to describe the
science that this amazing telescope will enable (Hook \cite{Hook05}),
not to mention the mass of science ideas that have been put together
by various instrument teams in developing Phase A studies and related
documents (from which I borrow quite extensively in this article).
There are also the documents that describe the science to be
undertaken with other future large telescopes (e.g.\ Silva et al.\
\cite{Silvaetal07}), as well as the cases that were developed for
projects that have now been dropped or merged with the current
programme (e.g.\ Andersen et al.\ \cite{Andersenetal03}), all of which
overlap significantly with the E-ELT case.  Second, even the most
optimistic timetable places first light for E-ELT eight years off.
One thing that we learned from the Hubble Space Telescope (HST) was
that the detailed science cases for such long timescale projects are
frequently overtaken by developments in the field long before they are
completed.  This is not to say that such science cases are not
important, as they demonstrate the kind of science that a telescope
can do, nor that there will be nothing left to do once the telescope
is completed -- as we have seen with HST, it is in the nature of
telescopes that they are remarkably flexible tools that can be turned
to a fascinating variety of tasks that were not envisaged when they
were designed.  Nonetheless, it does indicate that it is not worth
trying to be too definitive in a short review like this, so I will
instead concentrate on broader themes and examples that illustrate the
range of science that we will be able to undertake.

I also need to define a little more clearly what is meant by the
``local Universe,'' since this term clearly refers to different scales
if one is studying solar physics or galaxy morphology.  Perhaps the
most useful definition is that this phrase describes a large enough
volume of the Universe to contain a representative sample of the
object under study, but not so large that the finite speed of light
means that we start seeing cosmic evolution in the population.  Note
that there is no guarantee that such a range of distances exists: for
example, bright quasars are so rare that any decent-sized sample will
span a range of redshifts that starts to probe the cosmic evolution of
the quasar population.  Nonetheless, it makes a reasonable working
definition for the purposes of this review.  Indeed, it was this
definition that provided one of the principal science drivers for a
telescope as large as the E-ELT: if we are to use this instrument to
study the detailed properties of galaxies, we need to sample a region
that contains all the environments in which these objects typically
reside, including clusters.  The closest such reasonably-rich systems
are the Virgo Cluster and the Fornax Cluster, which motivated the
requirement for a telescope that can study resolved stellar
populations at distance of $\sim 20\,{\rm Mpc}$.

In the remainder of this paper, a number of examples of this kind of
local Universe science that will be enabled with the E-ELT are briefly
outlined, while Section~\ref{sec:conclusions} concludes with a
cautionary note about the risks of over-reliance on such cases.

\section{Young Starburst Clusters}
An excellent example of the ability of the E-ELT to allow us to study
a more representative sample of the Universe is provided by young
starburst clusters.  At present, the only two such clusters that can
be studied in any detail are Westerlund 1 and 2.  Even these
relatively nearby examples, at distances of only a few kiloparsecs,
are difficult to study, lying in the Galactic plane and hence being
obscured by visual extinctions of $A_V \sim 13$ and $5$ respectively
(Piatti, Bica \& Claria \cite{Piattietal98}).  With the E-ELT, we will
win over current technology in several ways.  First, the large
collecting area provides an effective mechanism for penetrating
extinction and seeing more distant clusters.  Second, the availability
of mid-infrared instrumentation will allow us to explore the
properties of these systems at wavelengths that are much less affected
by such extinction.  Third, the unprecedented spatial resolution will
allow us to resolve individual stars, even in much more distant
clusters, allowing us to study their individual properties.  In this
way, we will also be able to study starburst clusters in other parts
of spiral arms, such as NGC~3603 further out along the Sagittarius
Arm, and RSGC~1 much further in along the Scutum-Crux Arm (Davies et
al.\ \cite{Daviesetal07}).  Similar attention can also be focused on
the two starburst systems near the Galactic Centre, the Arches Cluster
and the Quintuplet Cluster.

Particular observations that could be made to study these systems
include measuring individual stellar proper motions to identify
cluster members unequivocally, resolving their stellar luminosity
functions all the way down to sub-stellar masses, and studying the
properties of the individual stars, such as searching for the
signatures of planetary disks from their mid-infrared spectra.  By
exploring examples all along the arms of a typical spiral galaxy like
the Milky Way as well as those forming near the centre of the Galaxy,
we will be able to search for any variations with environment, some of
which would have far-reaching consequences.  For example, if for some
reason planetary systems were only found to start to form in clusters
over a limited range of environments, that would have major
implications for the study and understanding of exo-planets.

\section{Evolved Stars}
At the other end of stellar evolution from the young starburst
clusters, the E-ELT will also shed new light on stars as they near the
ends of their lives.  Here, the issue is that this phase of stellar
evolution is short, so these objects are rare and the number
accessible to study with 8-metre telescopes is very limited.  Even
with the few nearby examples, the subtlety of the spatial structure in
dust emission as these objects lose their outer envelopes means that
very high resolution mid-infrared data are required to understand
their properties (e.g.\ Weigelt et al \cite{Weigeltetal98}).  With the
reach of the E-ELT, we will finally be in a position to study a
representative sample of these objects, quantify the range of
morphologies that they display as they shed their outer envelopes,
explore the spatially-resolved emission from a range of molecules, and
investigate how these properties vary with environment.  In other
words, for the first time we will be able to study them properly in
the context of the local Universe.

\section{The Galactic Centre}
Another new vista opened up by the E-ELT will be that of the Galactic
Centre.  Although we have been studying the properties and motions of
stars in this region for quite some time (Ghez et al.\
\cite{Ghezetal00}; Genzel et al.\ \cite{Genzeletal00}), we are really
only exploring the tip of the iceberg.  For example, there is some
evidence that the Galactic Centre region is not conducive to the
formation of lower-mass stars (Nayakshin \& Sunyaev
\cite{NayakshinSunyaev05}), but it will only be with the E-ELT that
we will be able to explore the luminosity function at very small radii
in the Galaxy all the way down to sub-stellar masses, to see quite how
important this effect really is.

The dynamics of these stars also offer an unprecedented probe of a
galactic nucleus: Weinberg, Milosavljevic \& Ghez
(\cite{Weinbergetal05}) showed that a telescope like the E-ELT will be
able to determine the distribution of mass in any central dark matter
cusp from its effect on the orbits of stars, as well as measuring the
mass of the central black hole with high precision.  By comparing the
proper motions of stars near the Galactic Centre to their
line-of-sight velocities inferred from spectra, Weinberg et al.\ also
showed that one will be able to determine the distance to the Galactic
Centre to a fraction of a percent.  They argue that such a measurement
could be used to remove the major uncertainty from current
measurements of the shape of the Galactic halo, which is a useful
diagnostic of the nature of dark matter (Olling \& Merrifield
\cite{OllingMerrifield00}).

Here we have to be a little careful, however, as while the Milky Way
may be a typical system, there is likely to be significant variance
from galaxy to galaxy in quantities such as central black hole mass
and halo flattening.  Thus, while such measurements may be of
interest, we are not really meeting the requirement on ``local
Universe'' science that we are exploring a representative sample of
objects, so ever more precise measurements may not translate into
ever more accurate scientific conclusions.

\section{The Fingerprints of Galactic Structure}\label{sec:finger}
Simulations have shown that the stellar halo of a galaxy like the
Milky Way is likely to be a messy tangle of stellar streams from
destroyed merging satellites (Bullock \& Johnston
\cite{BullockJohnston05}). Indeed, the brightest and youngest such
streams are now readily accessible to observation (Belokurov et al.\
\cite{Belokurovetal06}).  As time passes, these streams become wrapped
and tangled to a point where they are no longer detectable as
photometric enhancements, so we cannot dig far back into the formation
of the Milky Way halo from such data.  However, in the phase space of
positions and velocities, their coherent structure remains apparent
for much longer, particularly if there is some way to tag stars as
likely having a common origin.  Similarly, within the disk of the
Milky Way one might hope to determine which stars originated in the
same stellar nurseries long after they have spread out around the
Galaxy.  In this regard, it is potentially invaluable that stars
originating in the same stellar cluster seem to have identical
detailed chemical abundances, but that those abundances vary
significantly from cluster to cluster (de Silva et al.\
\cite{deSilvaetal07}).  Thus by obtaining high-resolution spectra of
Milky Way stars, one not only measures the line-of-sight velocity
component of the phase space coordinates, and spectral indices that
may provide an indication of properties such as the star's age and
metallicity, but one also obtains a ``fingerprint'' of the star that
will uniquely tie it to its siblings.  Ryde (\cite{Ryde10}) has
demonstrated that appropriate diagnostic lines exist in the near
infrared for the E-ELT to exploit in exactly this exercise.

Most of the existing data at reasonably large distances (such as the
Galactic bulge) has been obtained from red giant stars, in which
convective mixing of processed material will have erased at least some
of these identifying chemical tags.  It has therefore not proved
possible to use such data to obtain unequivocal properties of these
stars and their origins, and one would really want to obtain spectra
of main-sequence stars where the chemical signals are much clearer.
The collecting area of E-ELT will enable us to obtain the necessary
high-resolution spectra of main-sequence stars all across the Galaxy
to perform such studies.  However, we can already obtain a taste of
what is to come by using what might be called the ``$\mu$-ELT,'' the
microlensing extremely large telescope.  Bensby et al.\
(\cite{Bensbyetal10}) have been exploiting the microlensing of bulge
stars by foreground objects, which amplifies their brightness to a point
where high-resolution spectra of even intrinsically-faint main
sequence stars can be obtained using 8-metre telescopes.  This
analysis has shown that bulge main-sequence stars have tightly
correlated properties in their chemical abundances, similar to those
seen in thick-disk stars, which offers an important new clue to their
origins.  These data also show that such spectra can provide reliable
age diagnostics for stars near the main-sequence turn-off.  However,
at present the data are limited by the rate at which such rare
microlensing events occur; with E-ELT, we will be able to make such
measurements for many times the sample of 15 stars for which Bensby et
al.\ (\cite{Bensbyetal10}) have been able to obtain observations.

\section{Nucleo-Chronometry}
With the kind of very high-resolution, very high signal-to-noise ratio
spectroscopy that is only possible with the light collecting power of
the E-ELT, one can attempt to make novel measurements that are almost
impossible at the moment.  For example, determination of the
abundances of heavy elements like uranium have only been made for a
tiny number of stars, and even in those cases the systematic
uncertainties in the measurements are quite large (Frebel et al.\
\cite{Frebeletal07}).  The motivation to make such measurements is
that these heavy elements are frequently radioactive, so their
abundances decrease with time.  By measuring those abundances relative
to a non-decaying similar element, one obtains a completely different
innovative measure of stellar age.  This ``clock,'' unlike more
conventional measures like the main-sequence turn-off, depends only on
the basic physics of radioactive decay, so is a very robust
measurement.  Further, it is sufficiently slow running that it can be
applied reliably to very old stars, giving a strong and independent
lower bound on the age of the Universe (Frebel et al.\
\cite{Frebeletal07}).  Such observations with the E-ELT at very high
resolution will reduce the systematic errors on these subtle
measurements, while observing much larger samples than are currently
accessible will increase the age of the oldest star observed, putting
a tighter bound on the cosmologically-interesting age of the Universe.

\section{Interpreting the Spite Plateau}
A further cosmologically-motivated measurement that can be obtained
from high-resolution stellar spectra comes from a study of the lithium
abundances in stars.  Spite \& Spite (\cite{SpiteSpite82}) found that
metal-poor main-sequence turn-off stars in the Milky Way all seem to
have remarkably similar abundances of lithium.  The natural
interpretation of this common value is that it reflects the amount of
the element produced by primordial nucleosynthesis in the Big Bang,
but unfortunately the measured abundance is around a factor of five
lower than that predicted by current cosmology (Cybert, Fields \&
Olive\ \cite{CybertFieldsOlive08}).  So, either there is something
fundamentally astray in our understanding of the Big Bang, or some
process had systematically depleted Galactic levels of lithium before
even these metal-poor stars had formed.

One way to discriminate between these possibilities would be to make
similar measurements in another galaxy, to see whether they contain
the same level of lithium (as would be predicted if it is truly
primordial), or different abundances (if something in the more local
environment had suppressed the element before the metal-poor stars had
formed).  Unfortunately, this is a painfully difficult measurement to
make, since it requires high resolution spectra of weak lines in
intrinsically faint stars.  However, with the unprecedented collecting
area of the E-ELT, and picking on our nearest neighbour galaxy the
Sagittarius Dwarf, it becomes possible.  Even then, it will require
several nights of integration to obtain the requisite signal-to-noise
ratio for the faint spectral features produced by lithium, but with a
suitable multiplexing spectrograph to obtain a number of Sagittarius
Dwarf turn-off stars simultaneously, it is a viable proposition that
could potentially overturn our current picture of cosmology.

\section{Resolved Stellar Populations}
As mentioned in the Introduction, one of the factors driving the
specifications of the E-ELT was to be able to resolve individual stars all
the way out to the closest clusters of galaxies.  Once individual
stars have been detected, every astronomer's immediate instinct is to
place them on a colour--magnitude (CM) plot.  At galaxy cluster distances
with the old stellar populations typically found in cluster galaxies,
we will struggle to get to faint enough luminosities to detect many of
the features in the CM diagram, but even just the red giant branch
offers a useful diagnostic for the metallicity distribution in these
systems.  Once we start looking at the relative populations of the
asymptotic giant branch and different colours of sub-giant, each of
which varies differently with the age of the stellar population, we
can start to obtain at least a crude measure of the age distribution
in these galaxies as well.

For more nearby systems, where we can obtain photometry all the way
down to the main-sequence turn-off, we will be able to construct a
complete and reasonably accurate star formation history of a galaxy in
a manner that is currently only possible in very nearby systems (Cole
et al.\ \cite{Coleetal07}).  Such analysis of a galaxy's
creation is sometimes referred to as ``galactic archeology,'' but the
exciting aspect here is that the measurement is so direct that what we
will be doing is much more akin to reading a historical record rather
than attempting an archaeological reconstruction from fragmentary
evidence.  

\section{A Cautionary Note}\label{sec:conclusions}
This review has briefly outlined a number of the local-Universe
science programmes that could be undertaken with the E-ELT.  Clearly,
invaluable results would be obtained from such studies, but I must
admit that I find some of them a little disappointing in their scope.
They tend to be rather conservative in what they seek to do, simply
pushing the envelope of understanding from existing measurements into
new physical environments.  If we are going to spend a billion euros
on a project like this, we surely have to do even better to justify
the expenditure.  And, with a telescope that is as large a leap
forward as the E-ELT, we really should be able to do just that.  So,
my plea at the end of this review is that we should be more innovative
and ambitious in our ideas for this telescope.

Let me give one example of the kind of crazy idea that I think we
should be considering.  A long-running question in the study of
galaxies has been the details of their mass distributions, and how
that mass might be divided between luminous and dark components.  For
spiral galaxies, the rotation curve provides a straightforward
diagnostic, but studies of elliptical systems are more complex, with
dynamical measures facing the uncertainty of the orbits followed by
the tracer, while other techniques such a gravitational lensing and
measuring the properties of hot X-ray gas are not really suitable for
the more ubiquitous low-mass ellipticals.  An innovative alternative
was put forward by Stiavelli \& Setti (\cite{StiavelliSetti93}), who
pointed out that the light escaping from radius $r$ in an elliptical
galaxy would lose energy as it does so, resulting in a gravitational
redshift of
\begin{equation}
v_{\rm grav}(r) \sim -\Phi(r)/c, 
\end{equation}
thus offering a direct probe of the gravitational potential $\Phi(r)$,
and hence the mass distribution of the galaxy.  The signal involved is
quite small, and is further diluted by line-of-sight projection
effects, but none-the-less should produce systematic redshifts in the
mean velocities of stars at small projected radii.  Coggins
(\cite{Coggins03}) showed that such redshifts are essentially
undetectable in the integrated spectrum of an elliptical galaxy.
However, he also calculated that for a compact elliptical like M32 one
should obtain a net redshift of $\sim 3\,{\rm km}\,{\rm s}^{-1}$ per
decade in projected radius.  While such a signal is quite small
compared to the random velocities of individual stars, averaging
together a few thousand stars at each radius would quickly beat down
the uncertainty to a point where this shift in the mean velocity is
easily measured.  In fact, by observing individual stars and drawing
on the spectral finger-printing techniques of Sect.~\ref{sec:finger},
we can do even better by targeting those whose chemical abundances
indicate they are at intrinsically small radii (e.g.\ Baes et al.\
\cite{Baesetal07}), thus significantly reducing the diluting effects
of line-of-sight projection and further increasing the redshift
signal.  The E-ELT's combination of high spatial resolution to pick
out individual stars, large collecting area to obtain adequate numbers
of photons from each star, and multiplexing instruments to record high
resolution spectra from large numbers of stars, could allow this kind
of crazy idea to become a reality.

We have a good few years before the E-ELT becomes a reality, so let's
spend at least a few of them thinking such out-of-the-box thoughts
rather than just rehashing a safe agreeable science programme.

\end{document}